\documentclass[epj]{svjour}
\usepackage{graphics}
\usepackage{cite}
\usepackage[title]{appendix}
\begin{document}
\title{Birthing of a daughter vesicle in a model system for self-reproduction vesicles}
\author{Petch Khunpetch\thanks{\emph{Present address:} School of Physical Sciences, University of Chinese Academy of Sciences, Beijing 100049, China. \email{petch@ucas.ac.cn}}, Yuka Sakuma, Masayuki Imai, \and Toshihiro Kawakatsu
%
}                     
%
%
\institute{Department of Physics, Tohoku University, Aoba, Aramaki, Aoba-ku, Sendai 980-8578, Japan}
\date{Received: date / Revised version: date}
%
\abstract{
Sakuma and Imai [Phys. Rev. Lett. \textbf{107}, 198101 (2011)] established a temperature-controlled cyclic process for a model system of self-reproducing vesicles without feeding. The vesicle generates a smaller inclusion vesicle called ``daughter vesicle'' inside the original vesicle (we call this ``mother vesicle'') and then the daughter vesicle is expelled through a small pore on the mother vesicle. This self-reproducing process is called birthing. In the present study we present theoretical model on the birthing process of a single, rigid daughter vesicle through a pore. By using a simple geometric picture, we derive the free energy constituting the material properties of the bending, stretching and line tension moduli of the mother vesicle, as a function of the distance between the centers of the daughter and mother vesicles, and the size of the daughter vesicle. We see clearly the disappearance of the energy barrier by selecting appropriate moduli. The dynamics of the system is studied by employing the Onsager principle. The results indicate that translocation time decreases as the friction parameter decreases, or the initial size of the daughter vesicle decreases. 
\PACS{
      {PACS-key}{discribing text of that key}   \and
      {PACS-key}{discribing text of that key}
     } 
} 
\maketitle
\section{Introduction}
\label{sec:intro}
Vesicles assembled from the amphiphilic molecules are considered as the important step in the transition pathway from molecular assembly to cellular life~\cite{SI1}. A vesicle can be regarded as a container for a virtual cell called a protocell which is composed of three fundamental components \textit{i.e.} a metabolism, genes, and a container~\cite{L, RBCDKPS, DS}. Developing self-reproducing vesicles coupled with the metabolic system is a step forward to protocell~\cite{DD, W}. Previous successful studies in developing model systems for self-reproducing vesicles~\cite{HFS, TTS, TS, WWL} have shown that the molecules of membrane precursor turn to assemble an inclusion vesicle called daughter vesicle inside the mother vesicle with the help of catalyst. When the daughter vesicle grows to certain size, the daughter vesicle is then expelled through the small pore on the mother vesicle in the process called ``birthing''. 

\begin{figure}
\centering
\resizebox{0.5\textwidth}{!}{%
  \includegraphics{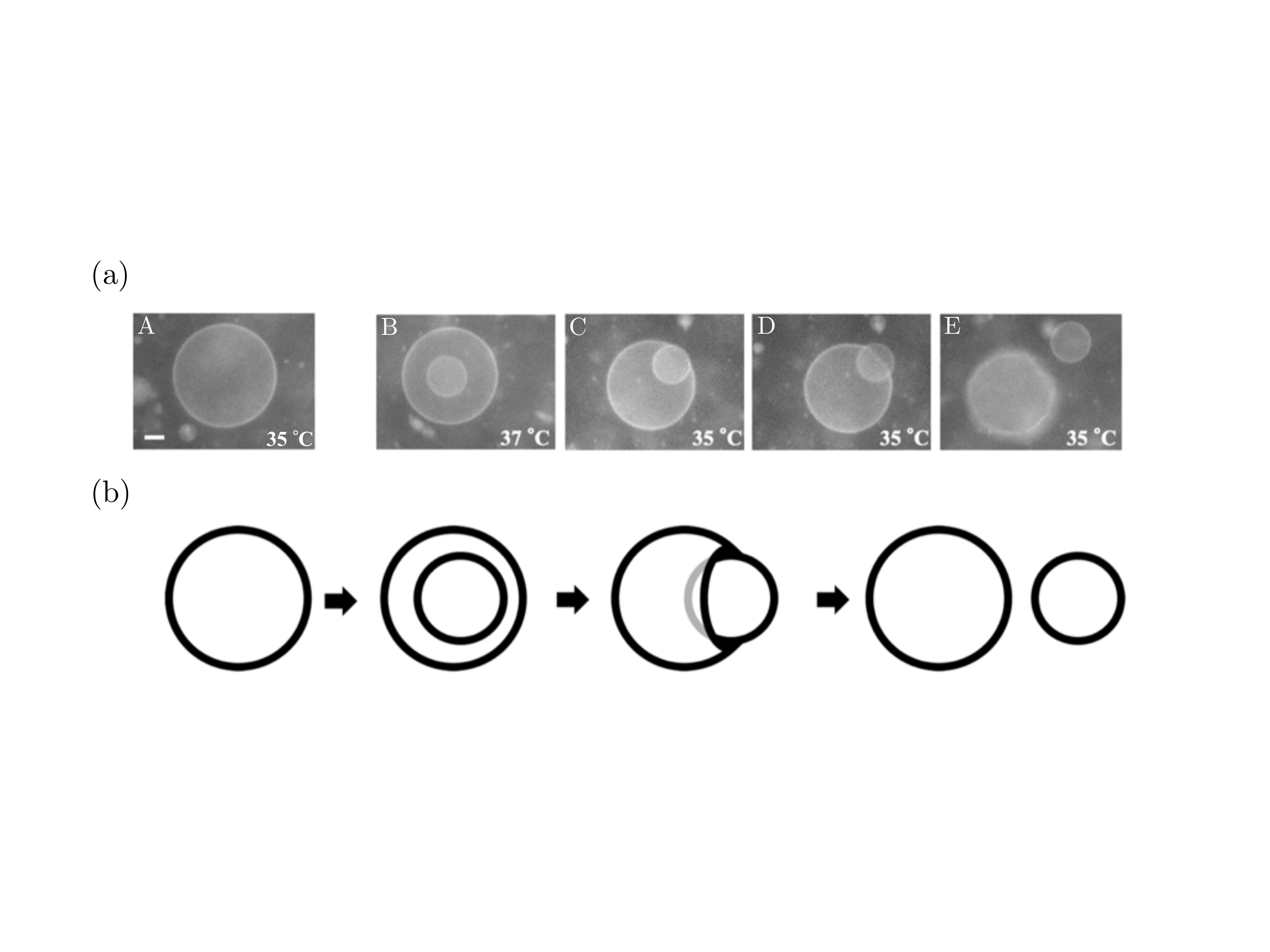}
}
\caption{(a) The spherical giant unilamellar vesicle (GUV) with a composition of DLPE/DPPC=3/7 is prepared at 35$^{\circ}$C (A). 
During the cooling process, the mother vesicle forms a pore by the surface tension due to the shrinkage of the membrane and the daughter vesicle is pushed into the pore due to the liquid flow coming out from the pore (B). Birthing of the daughter vesicle through a pore, where the temperature is maintained at 35$^{\circ}$C (C-E). About $90\%$ of around 50 mother GUVs with the diameter of $20-100\, \mu\textrm{m}$ exhibited the birthing where the averaged mother to daughter vesicle size ratio was about 0.4. The scale bar indicates $5\, \mu\textrm{m}$. (b) A schematic picture for birthing process in self-reproducing vesicles. (The micrographs and schematic representations adapted with permission from ref.~\cite{SI}. Copyrighted by the American Physical Society.)}
\label{fig:transition pathway}       
\end{figure}

Sakuma and Imai developed a model system for self-reproduction of giant unilamellar vesicle (GUV) without adding molecules~\cite{SI}. In their system, the vesicle with a composition of inverse-cone-shaped lipid having a small head group and bulky tails [1, 2-dilauroyl-\textit{sn}-glycero-3-phosphoethanolamine (DLPE)]/cylinder-shaped lipids [1, 2-dipalmitoyl-\textit{sn}-glycero-3-phosphocholine (DPPC)]=3/7 has a spherical shape at 35$^{\circ}$C. DPPC and DLPE have melting temperatures $T_{m}^{\textrm{{\scriptsize DPPC}}}=41^{\circ}$C and $T_{m}^{\textrm{{\scriptsize DLPE}}}=29 ^{\circ}$C, respectively. By increasing temperature from 35$^{\circ}$C to 42$^{\circ}$C, above $T_{m}^{\textrm{{\scriptsize DPPC}}}$, surface area of vesicle increases, and the GUV deforms to a stomatocyte shape. The stomatocyte vesicle then forms an inclusion vesicle inside it by plucking off the invagination neck. When the temperature is decreased to 35$^{\circ}$C, the surface area of the mother vesicle decreases. This causes an increase in the surface tension of the mother vesicle. To release this tension, the mother vesicle produces a pore and, then, the daughter vesicle is expelled through the pore. Driven by the line tension around the pore, the mother GUV recovers a spherical shape by closing the pore after the birthing. The observed pathway and the schematic diagram for birthing process in self-reproducing vesicles are shown in fig.~\ref{fig:transition pathway}. 

A coarse-grained description can give a reasonable pa-
rametrization of the complexity of specific chemical compounds that constitute the vesicles in terms of mechanical characteristics such as bending and stretching moduli. This approach is successful in the study of, for example, the problem of encapsulation of a rigid spherical particle by a vesicle~\cite{DG, MM}, and translocation of a vesicle through a narrow pore~\cite{SM, KMKD}. In the present study, we propose a theoretical model on the birthing process of a single, rigid daughter vesicle through a pore on the mother GUV. By using a simple geometric picture, we first derive the free energy within the framework of the Helfrich theory constituting the material properties of the mother vesicle, \textit{i.e.}, bending, stretching and line tension moduli, as a function of the distance between centers of the daughter and mother vesicles, and the size of the daughter vesicle. For studies of a closed vesicle, the contribution from the Gaussian curvature in the Helfrich free energy is just a constant and can then be omitted from the theory. However, when a pore is presented, the Gaussian curvature term cannot be neglected. We have evaluated the integral of the Gaussian curvature explicitly by using differential geometry of curved surface for the non-preserved topology of the mother vesicle. The theoretical study of pored vesicle is scarce. The stability of a budding pore with the introduction of the line tension energy is studied theoretically by Yao \textit{et al.}~\cite{YSTC}. The pored membrane is modelled as a spherical cap combined with a catenoid, where the boundary of a pore is located on the waist of the catenoid. The budding pore which has a meta-stable state is found but the meta-stable state is probably hard to observe experimentally because the energy barrier is very shallow and the range of the values of the line tension energy is very narrow. Experimental evidence that shows a pored vesicle is studied by Karatekin \textit{et al.}~\cite{KSGBPW}. By using visible light to increase the membrane tension, the vesicle is stretched until the membrane responds by opening a pore. The size of the pores are several microns but only a single pore can be observed at a time. As the inner liquid leaks out, the pores are transient which are driven to close by the line tension within a few seconds. Other experiment that shows the transient pores in GUVs has done by Rodriguez \textit{et al.}~\cite{RCP}. Sakuma's observation~\cite{YS} suggests that the time for resealing the pore when the daughter vesicle was detached is around a second. However, the pore lifetimes can be increased up to several minutes by adding some detergents. 

The dynamics of the system is studied by employing the Onsager principle, which is a variational principle originally proposed by L. Onsager~\cite{O1, O2}. The principle states that if a non-equilibrium system is described by a set of slow variables ${\bf x}=(x_{1}, x_{2}, ..., x_{f})$, the time evolution equation of the system can be obtained by minimization of Rayleighian defined by
\begin{equation}
\mathcal{R}=\Phi+\dot{F},
\end{equation}
with respect to ${\bf{\dot{x}} }\equiv\{\dot{x_{i}}\}$, where $\Phi$ is the energy dissipation function which is a quadratic form of the velocity and $\dot{F}$ is the time derivative of the free energy. The principle is a fundamental framework which is useful for deriving evolution equations in various soft matter systems~\cite{D0, D1, D2} such as gel dynamics equations~\cite{D2, D3}, Cahn-Hilliard equations of phase separation dynamics~\cite{O}, diffusion equations for particles in solutions~\cite{D0, D2} etc.

In our previous study, we employed the Onsager principle to study vesicle translocation through a narrow pore in a solid membrane separating two chambers~\cite{KMKD}. A similar work using the Fokker-Planck formalism has been done by Shojaei \textit{et al.}~\cite{SM}. In both works, the shape of the translocating vesicle changes due to the material constituting the wall which is much harder than the vesicle membrane. In the present work, the material constituting the mother and daughter vesicles are identical. As the shape of the daughter vesicle does not change considerably during the birthing process as seen in fig.~\ref{fig:transition pathway} (a), we then assume that the daughter vesicle size is fixed. Consequently, there is only one slow variable which is the distance between centers of the daughter and mother vesicles. The free energy of the system is derived by accounting for the bending and the stretching contributions from the mother vesicle. In order to understand the pore formation, the line tension energy is introduced. By selecting the appropriate bending, stretching, and line tension moduli, we see clearly the disappearance of the energy barrier which means the successful birthing process. The dynamics of the system indicates that the birthing (translocation) time decreases as the friction against the moving daughter vesicle due to the mother vesicle decreases, or the initial size of the daughter vesicle decreases.

This paper is organized as follows. In sect.~\ref{sec:model}, we describe the model for the system. Section~\ref{sec:results and discussions} is devoted to the discussion of the results of the birthing process. Finally, the conclusion of this study is given in sect.~\ref{sec:conclusions}.

\section{Theory of the birthing process}
 \label{sec:model}
 
  \subsection{\label{subsec:Geometry}Geometry}
  
\begin{figure}
\centering
\resizebox{0.7\textwidth}{!}{%
  \includegraphics{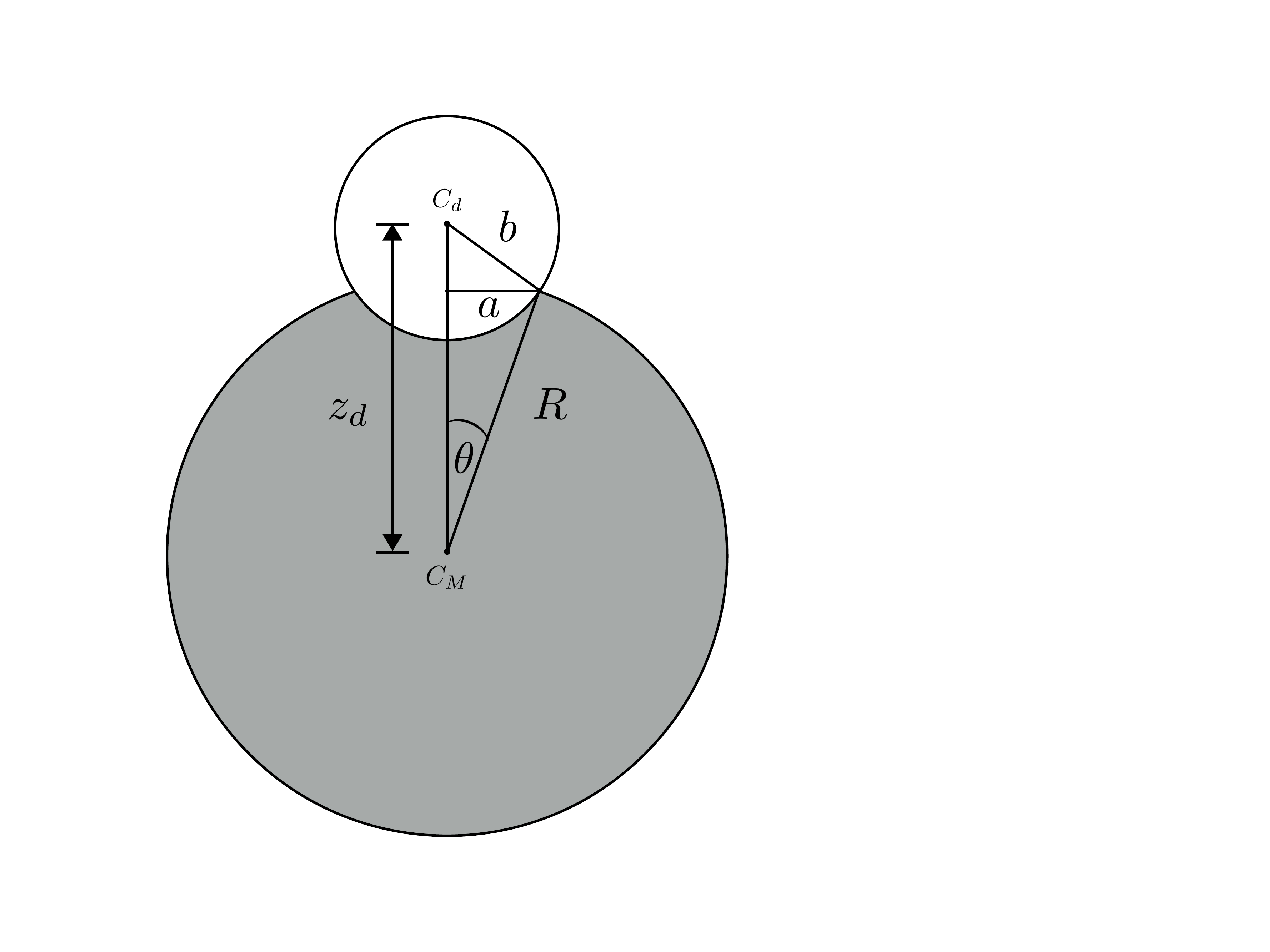}
}
\caption{Model of a birthing daughter vesicle in the intermediate state of the birthing process. The radius of the rigid daughter vesicle is $b$. Shaded region shows the volume between the mother and daughter vesicles, which is conserved during the birthing process. The radius of the circular pore is denoted by $a$.}
\label{fig:birthing model}
\end{figure}

The birthing of a single, spherical, rigid daughter vesicle with radius $b$ through a pore of the spherical mother vesicle with radius $R$ is modeled by a simple geometric ansatz as illustrated in fig.~\ref{fig:birthing model}. 
We neglect the thickness of the membrane of the mother vesicle. The distance between the center of the mother vesicle $(C_{M})$ and that of the daughter vesicle $(C_{d})$ is denoted by $z_{d}$. During the birthing process, the pore on the surface of the mother vesicle gradually expands and then shrinks due to the line tension energy. We have assumed that the pore has a circular shape with radius $a$ where its center is located on the straight line connecting the centers of the mother and daughter vesicles. We introduce the angle $\theta$ defined by the line joining the centers of the mother and daughter vesicles, and the line between the center of the mother vesicle and the edge of the hole (see fig.~\ref{fig:birthing model}). We assume that during the birthing process there is no liquid leaks out from the mother vesicle and that the daughter vesicle moves very slowly. The volume between the mother and daughter vesicles, which is indicated with the shaded region, is conserved. The membrane of the mother vesicle is assumed to have a surface tension and, in the initial state, the daughter vesicle is a little bit expelled from the mother vesicle which has radius $R_{0}$. In our assumption, the daughter vesicle is discharged due to the surface tension of  the membrane of the mother vesicle. The distance and the angle at the initial state are denoted by $z_{d,0}$ and $\theta_{0}$, respectively. When the daughter vesicle is entirely expelled from the mother vesicle and the pore is completely closed, we will have $\theta =0$. Then, $\theta$ has a maximum value when the daughter vesicle passes the mother vesicle with almost half of its volume. Investigation of fig.~\ref{fig:birthing model} 
shows that the three variables $R, z_{d},$ and $\theta$ are related with each other by the following geometrical constraint
\begin{equation}
\label{cosine law}
\cos\theta=\frac{R^{2}+z_{d}^{2}-b^{2}}{2z_{d}R}.
\end{equation}
It follows that we may take $z_{d}$ as the only one independent variable that describes the system. Since, the amount of the liquid in the volume between the two vesicles is preserved, it leads to another constraint as
\begin{eqnarray}
\label{volume conservation}
\Delta V&=&\Bigg\{\frac{4}{3}\pi R^{3}-\pi b^{2}\Big[(R\cos\theta-z_{d})+b\Big]\nonumber\\
&&+\frac{\pi}{3}\Big[(R\cos\theta-z_{d})^{3}+b^{3}\Big]\nonumber\\
&&-\pi R^{3}(1-\cos\theta)+\frac{\pi R^{3}}{3}(1-\cos^{3}\theta)\Bigg\}\nonumber\\
&&-\Bigg\{\frac{4}{3}\pi R_{0}^{3}-\pi b^{2}\Big[(R_{0}\cos\theta_{0}-z_{d,0})+b\Big]\nonumber\\
&&+\frac{\pi}{3}\Big[(R_{0}\cos\theta_{0}-z_{d,0})^{3}+b^{3}\Big]\nonumber\\
&&-\pi R_{0}^{3}(1-\cos\theta_{0})+\frac{\pi R_{0}^{3}}{3}(1-\cos^{3}\theta_{0})\Bigg\}\nonumber\\
&=0,
\end{eqnarray}
where the first and the second parentheses account for the present and the initial volume of the liquid.

\subsection{\label{subsec:free energy}The free energy}

We have assumed that, at equilibrium, the mother vesicle has a spherical shape. The total free energy of the system is given by
\begin{eqnarray}
\label{free energy}
F&=&\Bigg[\frac{\kappa}{2}\int(2H-c_{0})^{2}dA+\kappa_{G}\int KdA\Bigg]\nonumber\\
&&+\frac{\lambda}{2}\frac{(A-A_{\textrm{eq}})^{2}}{A_{\textrm{eq}}}+\sigma\oint dl.
\end{eqnarray}

The first two terms on the right hand side of eq.~(\ref{free energy}) are the bending energy which is harmonic in the mean curvature $H$, where $H$ is defined by $H\equiv(1/R_{1}+1/R_{2})/2$ ($R_{1}$ and $R_{2}$ stand for the radii of principal curvatures.) and $c_{0}$ is the spontaneous curvature, which is zero when there is no asymmetry between the inner and outer leaflets of the bilayer. Since the mother vesicle does not preserve its topology, the Gaussian curvature, $K\equiv1/(R_{1}R_{2})$, needs to be taken into account. $\kappa$ and $\kappa_{G}$ are the local bending and the Gaussian bending rigidity, respectively. Typically, the bending modulus $\kappa$ is of the order of 10 $k_{B}T$ for fluid membranes and the experimental techniques to measure this bending modulus and its reported values are given in ref.~\cite{WH, ROMNE, WP, M}. The Gaussian bending modulus has a negative sign which indicates that the membrane prefers the surface with lower genus. A few results of the ratio $(-\kappa_{G}/\kappa)$ for bilayers are available in ref.~\cite{HBD}. For example, the ratio is $0.9\pm0.38$ for the various mixing ratios of [dioleoyl-\textit{sn}-glycero-3-phosphatidylcholine (DOPC)]/[sphingomyelin (S
M)]/[cholesterol (Chol)]~\cite{BDWJ}. In this study, we have simply selected $\kappa_{G}/\kappa=-1$. According to the Gauss-Bonnet theorem~\cite{S}, the integral of the Gaussian curvature yields
\begin{equation}
\label{Gauss-Bonnet}
\int KdA=4\pi(1-g),
\end{equation}
where the number $g$ is the genus of the surface, for example $g=0$ is a sphere, while a torus has $g=1$. The integrations are $4\pi$ (for a sphere) and $0$ (for a torus) where the spherical shape is more stable than a torus (without the consideration of the integral of the mean curvature). It is also possible to evaluate $\int KdA$ without using the Gauss-Bonnet theorem. For example, since the Gaussian curvature $K$ for the sphere with radius $R$ is $1/R^{2}$ and the area $A=4\pi R^{2}$, then, the integral of the Gaussian curvature simply yields~\cite{G}
\begin{equation}
\int KdA=K\cdot A=\frac{1}{R^{2}}\cdot(4\pi R^{2})=4\pi.
\end{equation}

The third term on the right hand side of eq.~(\ref{free energy}) shows the stretching energy of the mother vesicle. It is in the harmonic form in the change of vesicle surface area $\Delta A=A-A_{\textrm{eq}}$, where $A_{\textrm{eq}}=4\pi R_{\textrm{eq}}^{2}$ is the equilibrium surface area of the mother vesicle and $A$ is the mother vesicle surface area subtracted by the area of the pore, which is determined by $R$ and $\theta$. The value of the stretching modulus $\lambda$ is of the order of $10^{8}\, k_{B}T/\mu\textrm{m}^{2}$~\cite{ER, PUWS}.

The last term on the right hand side of eq.~(\ref{free energy}) refers to the line tension energy of the perimeter of the circular pore. Karatekin \textit{et al.}~\cite{KSGBPW} reported line tension modulus $\sigma$ is of the order of $10^{3}-10^{4}\, k_{B}T/\mu\textrm{m}$ for bare DOPC bilayers. 

We can analytically calculate the bending, stretching, and line tension energies of the system based on the differential geometry of a curved surface. First, let us calculate the contribution to the free energy from the bending. The Helfrich free energy is given by
\begin{equation}
\label{bending}
F_{b}=\frac{\kappa}{2}\int(2H-c_{0})^{2}dA+\kappa_{G}\int KdA.
\end{equation}
The integral $\int dA$ is taken over the area of the mother vesicle in which the area of the pore is eliminated, \textit{i.e.}
\begin{eqnarray}
\label{surface area}
A&=&A_{\textrm{mother vesicle}}-A_{\textrm{pore}}\nonumber\\
&=&2\pi R^{2}(1+\cos\theta).
\end{eqnarray}
Then, the bending energy can be written as
\begin{eqnarray}
\label{bending}
F_{b}&=&\pi\kappa R^{2}(1+\cos\theta)\Bigg(\frac{2}{R}-c_{0}\Bigg)^{2}+2\pi\kappa_{G}(1+\cos\theta)\nonumber\\
&=&\pi\kappa(1+\cos\theta)\Bigg(R^{2}\Bigg(\frac{2}{R}-c_{0}\Bigg)^{2}+\Bigg(\frac{2\kappa_{G}}{\kappa}\Bigg)\Bigg),
\end{eqnarray}  
where the mean curvature $H$ and the Gaussian curvature $K$ for the sphere with radius $R$ are $1/R$ and $1/R^{2}$, respectively and the spontaneous curvature is $c_{0}.$ Using eq.~(\ref{cosine law}), we can eliminate $\cos\theta$. Then, we obtain the Helfrich free energy in terms of the parameters $R, z_{d},$ and $b$ as
\begin{equation}
F_{b}=\pi\kappa\Bigg(1+\frac{R^{2}+z_{d}^{2}-b^{2}}{2z_{d}R}\Bigg)\Bigg(R^{2}\Bigg(\frac{2}{R}-c_{0}\Bigg)^{2}+\Bigg(\frac{2\kappa_{G}}{\kappa}\Bigg)\Bigg).
\end{equation}

In order to obtain the stretching contribution, we must calculate the surface area of the mother vesicle subtracted by the area of the circular pore which is already given by eq.~(\ref{surface area}).
Then, the stretching energy can be written as
\begin{eqnarray}
\label{stretching}
F_{s}&=&\frac{\lambda}{2}\frac{(\Delta A)^{2}}{A_{\textrm{eq}}}\nonumber\\
&=&\frac{\pi\lambda}{2R_{\textrm{eq}}^{2}}[R^{2}(1+\cos\theta)-2R_{\textrm{eq}}^{2}]^{2}\nonumber\\
&=&\frac{\pi\lambda}{2R_{\textrm{eq}}^{2}}\Bigg[R^{2}\Bigg(1+\frac{R^{2}+z_{d}^{2}-b^{2}}{2z_{d}R}\Bigg)-2R_{\textrm{eq}}^{2}\Bigg]^{2},
\end{eqnarray}
where the law of cosine eq.~(\ref{cosine law}) comes to the end for eliminating $\theta$.

Likewise, we calculate the line tension energy. It is given by
\begin{equation}
\label{line tension}
F_{l}=2\pi\sigma R\sin\theta.
\end{equation}
Since $\cos\theta=(R^{2}+z_{d}^{2}-b^{2})/(2z_{d}R)$, the identity $\sin^{2}\theta+\cos^{2}\theta=1$ gives the relation
\begin{equation}
\sin\theta=\sqrt{1-\frac{(R^{2}+z_{d}^{2}-b^{2})^{2}}{(2z_{d}R)^{2}}}.
\end{equation}
Then, the line tension energy becomes
\begin{equation}
F_{l}=2\pi\sigma R\sqrt{1-\frac{(R^{2}+z_{d}^{2}-b^{2})^{2}}{(2z_{d}R)^{2}}}.
\end{equation}

Finally, the total free energy of the system as a function of the radius $R$, the distance $z_{d}$, and the radius of the daughter vesicle $b$ is given by
\begin{eqnarray}
\label{full free energy}
F[R,z_{d};b]&=&\pi\kappa\Bigg(1+\frac{R^{2}+z_{d}^{2}-b^{2}}{2z_{d}R}\Bigg)\nonumber\\
& &\times\Bigg(R^{2}\Bigg(\frac{2}{R}-c_{0}\Bigg)^{2}+\Bigg(\frac{2\kappa_{G}}{\kappa}\Bigg)\Bigg)\nonumber\\
& &+\frac{\pi\lambda}{2R_{\textrm{eq}}^{2}}\Bigg[R^{2}\Bigg(1+\frac{R^{2}+z_{d}^{2}-b^{2}}{2z_{d}R}\Bigg)-2R_{\textrm{eq}}^{2}\Bigg]^{2}\nonumber\\
& &+2\pi\sigma R\sqrt{1-\frac{(R^{2}+z_{d}^{2}-b^{2})^{2}}{(2z_{d}R)^{2}}}.
\end{eqnarray}
Here we should note that $R$ is not independent of $z_{d}$. As we have assumed that during the birthing process the volume of the region between the mother and the daughter vesicles shown in the shaded region in fig.~\ref{fig:birthing model} is preserved. Then, $R$ can be determined by the volume constraint eq.~(\ref{volume conservation}) together with geometric constraint eq.~(\ref{cosine law}). Finally, the free energy eq.~(\ref{full free energy}) is a function of $z_{d}$ and $b$ only.

\subsection{The energy dissipation function}
 \label{subsec:dissipation function}
 
When the daughter vesicle is moving through the pore, the energy dissipation is caused mostly at the pore. In the limit of small Reynolds number, the dissipation function for the daughter vesicle is given by
\begin{equation}  
\label{dissipation vesicle}
\Phi=\frac{1}{2}\zeta \dot{z_{d}}^{2},
\end{equation}
where $\zeta$ is the friction coefficient. Generally, the friction coefficient is a function of the slow variables and it can be derived from the Stokesian hydrodynamics. In this work, for simplicity, we assume that the friction coefficient is a constant which is proportional to the radius of the pore $a$. As we assume that the daughter vesicle is a rigid body, there is no liquid flow inside of the daughter vesicle. Thus, the friction is generated at the perimeter of the pore only and can be written as 
\begin{eqnarray}  
\zeta&=&2\pi\alpha a\nonumber\\
&=&2\pi\alpha R\sin\theta,
\end{eqnarray}
where $\alpha$ is related to the dynamic viscosity and has the unit of kg m$^{-1}$ s$^{-1}$. The estimated order of magnitude of $\alpha$ for $b=10\, \mu\textrm{m}$, the translocation time of the daughter vesicle $t_{\textrm{trans}}\approx1.0\, \textrm{s}$, and $\lambda=1.0\times 10^{8}\, k_{B}T/\mu\textrm{m}^{2}$, is around $10^{3}\,$ kg m$^{-1}$ s$^{-1}$ (see Appendix A.). Please note that the dynamic viscosity of water is around $8.90\times 10^{-4}\,$ kg m$^{-1}$ s$^{-1}$ at $25^{\circ}$C~\cite{R}, while, in our assumption, there is no water layer between the daughter vesicle and the perimeter of the mother vesicle's pore.  
 
\subsection{The equation of motion}
 \label{subsec:EOM}
 
The Rayleighian of the system is given by
\begin{eqnarray}  
\mathcal{R}&=&\Phi+\dot{F}\nonumber\\
&=&\frac{1}{2}\zeta \dot{z_{d}}^{2}+\frac{dF(z_{d})}{dt}\nonumber\\
&=&\frac{1}{2}\zeta \dot{z_{d}}^{2}+\frac{dF}{dz_{d}}\dot{z_{d}}.
\end{eqnarray}

By minimizing $\mathcal{R}$ with respect to $\dot{z_{d}}$, \textit{i.e.}, $\partial\mathcal{R}/\partial\dot{z_{d}}=0$, we obtain the equation of motion for the daughter vesicle
\begin{equation}  
\label{equation_birthing}
\dot{z_{d}}=-\frac{1}{\zeta}\frac{dF}{dz_{d}}.
\end{equation}

\section{Results and discussions}
 \label{sec:results and discussions}
 
We present our results by first discussing the explicit free energy landscapes when the bending, stretching, and line tension moduli are changed, while the daughter vesicle size is constant. Later, the kinetics of birthing will be discussed when the friction coefficient is changed or the size of the daughter vesicle is varied.  

\subsection{Free energy landscapes}
\label{subsec:energy landscapes}

\begin{figure}
\centering
\resizebox{0.71\textwidth}{!}{%
  \includegraphics{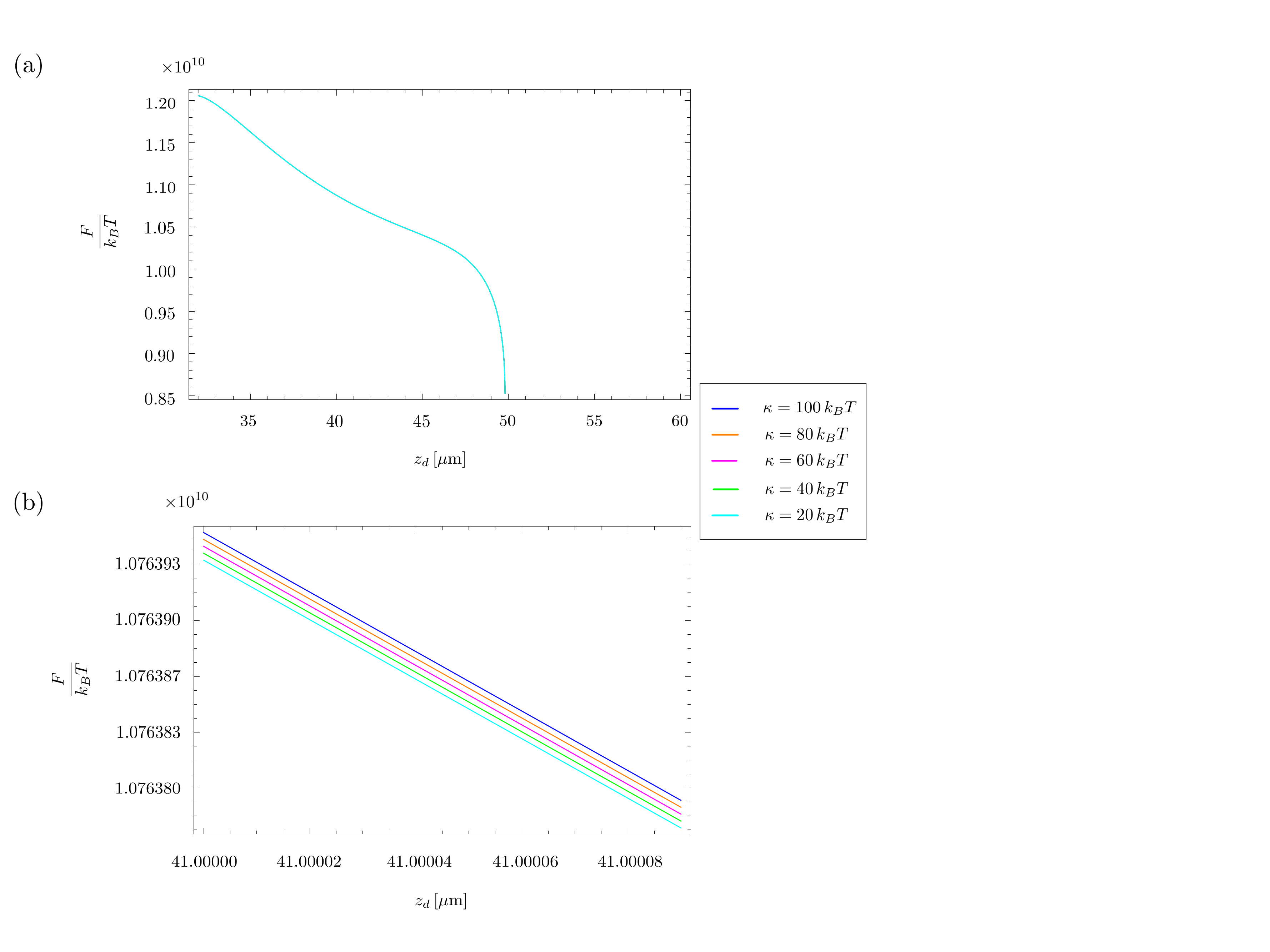}
}
\caption{The change in the free energy landscapes when the bending modulus $\kappa$ changes. The stretching and line tension moduli are fixed at $\lambda=1.0\times 10^{8}\,  k_{B}T/\mu\textrm{m}^{2}$ and $\sigma=0.7\times 10^{8}\,  k_{B}T/\mu\textrm{m}$, respectively. The $\kappa/\kappa_{G}$ ratio is $-1$. The size of the daughter vesicle is $b=10\, \mu\textrm{m}$, while the equilibrium size of the mother vesicle is $R_{\textrm{eq}}=38\, \mu\textrm{m}$. The initial size of the mother vesicle is $R_{0}=40\, \mu\textrm{m}$. The spontaneous curvature $c_{0}$ is set as zero. The plots are shown for the entire landscapes (a) and at very small scale (b).}
\label{fig:energy_bending}
\end{figure}

\begin{figure}
\centering
\resizebox{0.715\textwidth}{!}{%
  \includegraphics{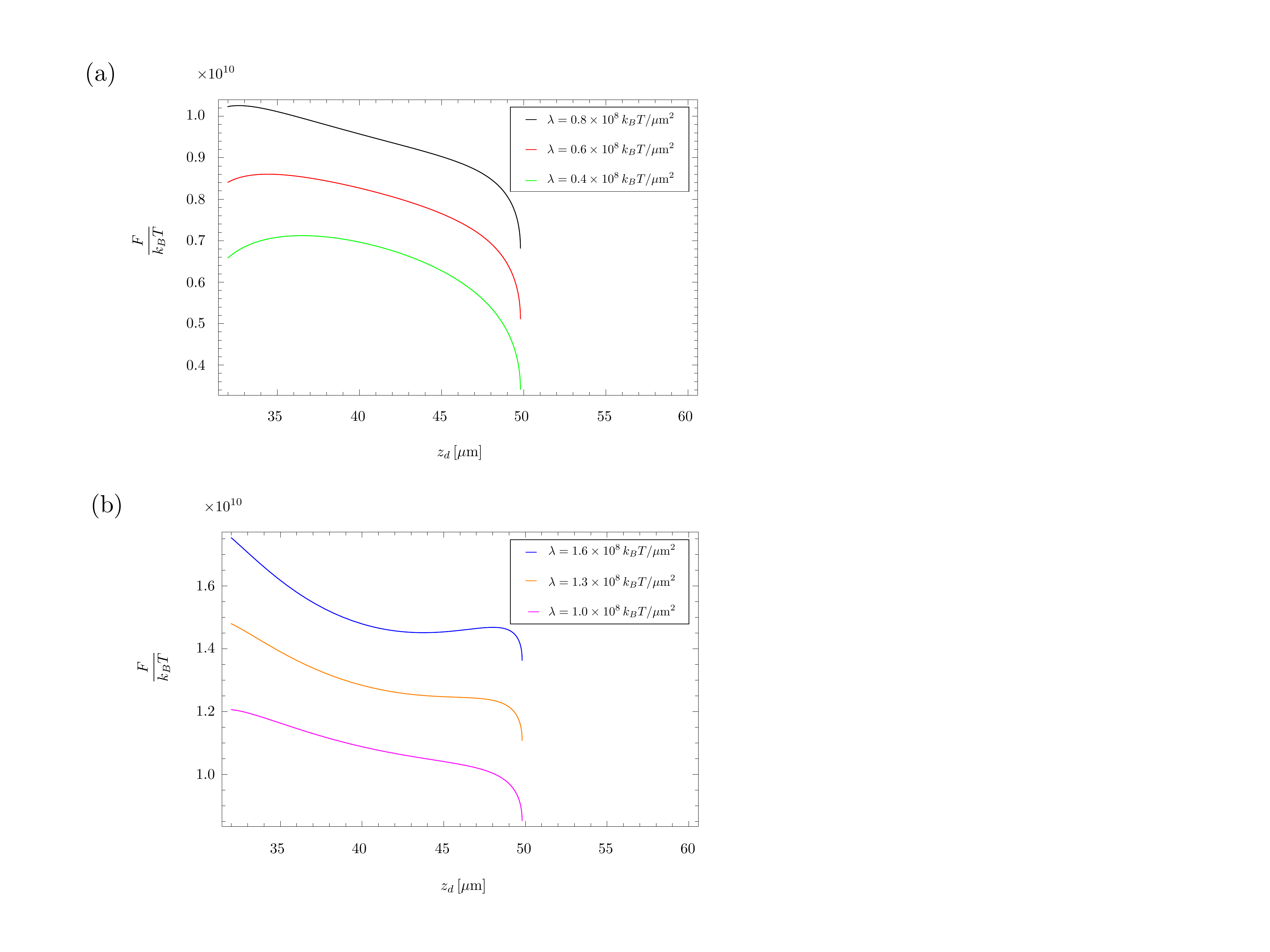}
}
\caption{The free energy landscapes with the variation of the stretching modulus $\lambda$. The bending and line tension moduli are fixed at $\kappa=20\, k_{B}T$, $\kappa_{G}=-20\, k_{B}T$, and $\sigma=0.7\times 10^{8}\,  k_{B}T/\mu\textrm{m}$, respectively. The equilibrium size of the mother vesicle is $R_{\textrm{eq}}=38\, \mu\textrm{m}$. The initial size of the mother vesicle is $R_{0}=40\, \mu\textrm{m}$. The size of the daughter vesicle is $b=10\, \mu\textrm{m}$. The initial distance that separates the mother and the daughter is $z_{d,0}=32\, \mu\textrm{m}$. The spontaneous curvature $c_{0}=0$.}
\label{fig:energy_stretching}
\end{figure}

\begin{figure}
\centering
\resizebox{0.715\textwidth}{!}{%
  \includegraphics{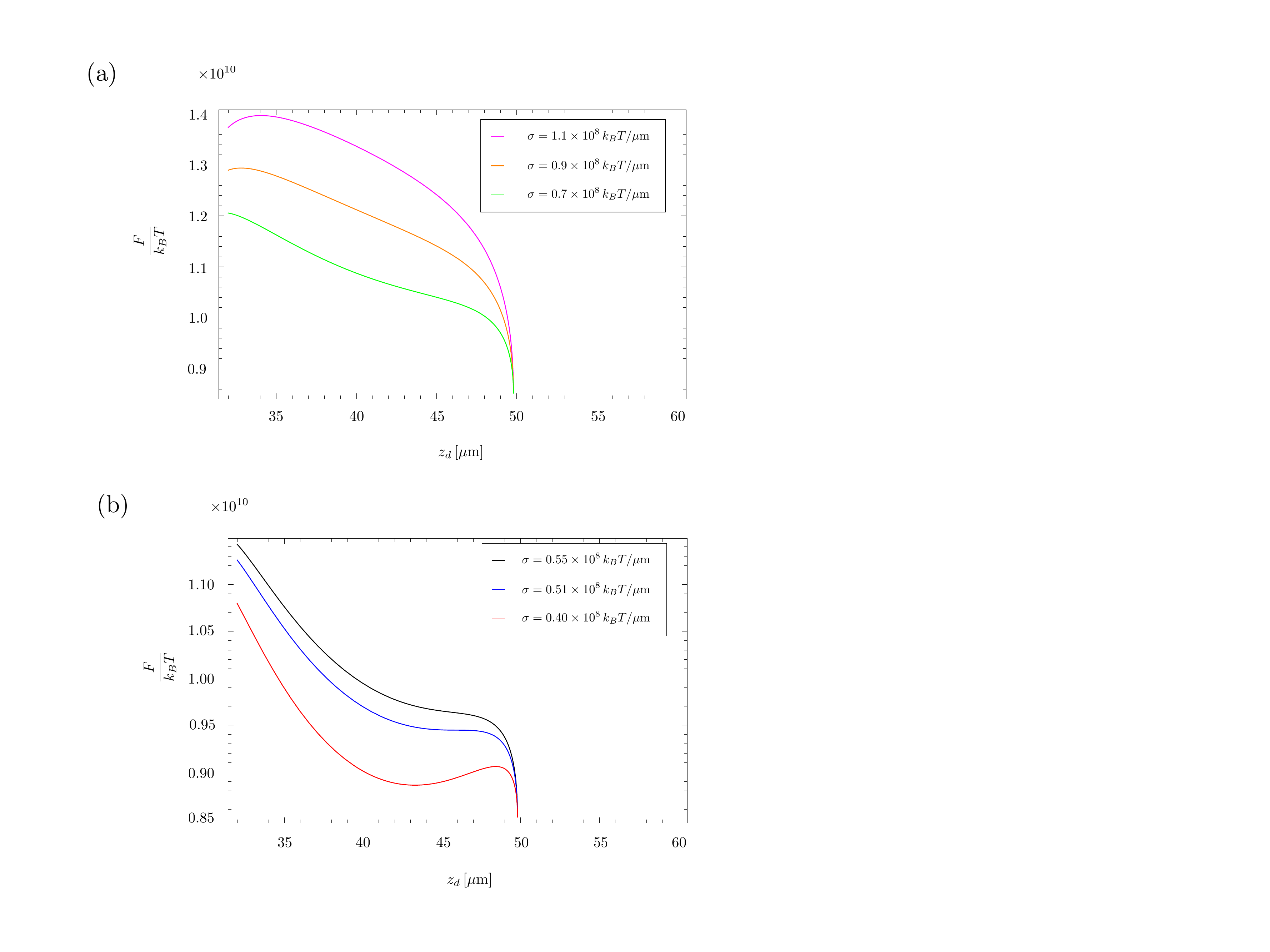}
}
\caption{The free energy landscape as a function of $z_{d}$ for each value of the line tension modulus $\sigma$. The local bending, Gaussian bending, and stretching moduli are fixed at $\kappa=20\, k_{B}T$, $\kappa_{G}=-20\, k_{B}T$, and $\lambda=1.0\times 10^{8}\,  k_{B}T/\mu\textrm{m}^{2}$, respectively. $b$ is fixed at $10\, \mu\textrm{m}$. $R_{\textrm{eq}}=38\, \mu\textrm{m}$, $R_{0}=40\, \mu\textrm{m}$, and $c_{0}=0$.}
\label{fig:energy_line}
\end{figure}

Representative results of the free energy landscapes over $z_{d}$ when the bending, stretching, and line tension moduli change are shown in figs.~\ref{fig:energy_bending},~\ref{fig:energy_stretching}, and ~\ref{fig:energy_line}.

All plots in figs.~\ref{fig:energy_bending}, ~\ref{fig:energy_stretching}, and~\ref{fig:energy_line} are evaluated when the size of the daughter vesicle is fixed at $b=10\, \mu\textrm{m}$, while the equilibrium size of the mother vesicle is taken to be $R_{\textrm{eq}}=38\, \mu\textrm{m}$. The initial size of the mother vesicle is $R_{0}=40\, \mu\textrm{m}$. The initial size means the state that the daughter vesicle is slightly expelled from the mother vesicle, where the pore has already been opened. When the daughter vesicle is discharged from the mother vesicle completely, the pore shrinks and eventually closes, and the mother vesicle recovers its spherical shape.

Figure~\ref{fig:energy_bending}(a) shows that all curves for each fixed value of $\kappa$ almost coincide, where the stretching and line tension moduli are fixed at $\lambda=1.0\times 10^{8}\, k_{B}T/\mu\textrm{m}^{2}$ and $\sigma=0.7\times 10^{8}\,  k_{B}T/\mu\textrm{m}$, respectively, and $\kappa/\kappa_{G}=-1$. Increasing of the free energy when the bending modulus $\kappa$ increases can be seen in fig.~\ref{fig:energy_bending}(b) at very small scale. All graphs show that there is no energy barrier and the daughter vesicle succeeds in the birthing process at $z_{d}<50\, \mu\textrm{m}$. As the initial size of the mother vesicle is $40\, \mu\textrm{m}$ and the size of the daughter vesicle is $10\, \mu\textrm{m}$, we have set up the initial distance that separates the mother and the daughter as $z_{d,0}=32\, \mu\textrm{m}$ and we found that the final value of $z_{d}$ is less than $50\, \mu\textrm{m}$. This means that, at the equilibrium final state, the mother vesicle reduces its size due to the stretching energy. From the result, it seems that changes in bending modulus does not affect the free energy landscapes significantly. Theoretical study done by Brochard-Wyart \textit{et al.}~\cite{BWGS} and experimental works of Karatekin \textit{et al.}~\cite{KSGBPW} and Rodriguez \textit{et al.}~\cite{RCP} for pored vesicles have suggested that the transient pores observed in stretched vesicles are driven by the competition between the surface tension and the line tension that our results would confirm their studies as we will show in the following.

In contrast to changes in the bending modulus, as we discussed above, figs.~\ref{fig:energy_stretching}(a) and (b) show the effect of varying the stretching modulus $\lambda$ on the free energy landscape. In this figure, the bending and line tension moduli are fixed at $\kappa=20\, k_{B}T$, $\kappa_{G}=-20\, k_{B}T$, and $\sigma=0.7\times 10^{8}\,  k_{B}T/\mu\textrm{m}$, respectively. In the range of $z_{d}$ where the tangent to the plot of $F/k_{B}T$ has a positive slope, the system shows the energy barrier. Increasing $\lambda$ will reduce the height of the barrier where the maximum point moves more and more closer to $z_{d,0}$ (fig.~\ref{fig:energy_stretching}(a)). The barrier disappears and the whole range of $z_{d}$ has a negative slope which means the daughter vesicle is discharged completely at $\lambda=1.0\times 10^{8}\, k_{B}T/\mu\textrm{m}^{2}$. However, increasing $\lambda$ further, the system shows the metastable state where the daughter vesicle is trapped in its vicinity as we can see in the curve of $\lambda=1.6\times 10^{8}\, k_{B}T/\mu\textrm{m}^{2}$ (fig.~\ref{fig:energy_stretching}(b)).  

The free energy landscapes when the line tension modulus $\sigma$ changes are shown in figs.~\ref{fig:energy_line}(a) and (b). The bending and stretching moduli are fixed at $\kappa=20\, k_{B}T$, $\kappa_{G}=-20\, k_{B}T$, and $\lambda=1.0\times 10^{8}\,  k_{B}T/\mu\textrm{m}^{2}$, respectively. We see clearly the decreasing in the energy barrier when $\sigma$ decreases. The daughter vesicle has succeeded in the birthing at $\sigma=0.7\times 10^{8}\, k_{B}T/\mu\textrm{m}$ when the energy barrier disappears (fig.~\ref{fig:energy_line}(a)). Further reducing $\sigma$, the plot of $F/k_{B}T$ has a stagnant point at $z_{d}=45.95\, \mu\textrm{m}$ on the curve of $\sigma=0.51\times 10^{8}\,  k_{B}T/\mu\textrm{m}$ and has a negative slope everywhere else. Continuously reducing $\sigma$, the plot of the free energy has a dip where the bottom of the dip corresponding to the metastable state is at $z_{d}=43.29\, \mu\textrm{m}$, as seen clearly in the curve of $\sigma=0.40\times 10^{8}\,  k_{B}T/\mu\textrm{m}$ (fig.~\ref{fig:energy_line}(b)). 

Further investigating the metastable states as seen in figs.~\ref{fig:energy_stretching}(b) and~\ref{fig:energy_line}(b) suggests that their origin is from the competition between the stretching energy and the line tension energy. The metastable state refers the state of the system other than the state with the least energy, where the daughter vesicle is trapped in the basin of this state. The stretching free energy penalty which is assumed to be the quadratic form in the change of the mother vesicle's surface area can be found to be minimized when the surface area of the mother vesicle subtracted by the area of the pore is minimum. While the line tension energy that attempts to decrease the edge length of the pore reaches the maximum value where the pore size opens maximally, by definition, and the daughter vesicle passes through the pore almost half of its volume. As we have already seen that, in our model, the bending energy did not play a significant role, the competition between the two energies gives rise the metastable state. 

Experimentally, it is rather difficult to directly measure the line tension modulus $\sigma$ of the membrane due to the instability of a pore. The reported value is rare. Our simulations suggest that $\sigma$ is to be of around the order of $10^{8}\, k_{B}T/\mu\textrm{m}$ for the successful birthing process with $\kappa$, $\kappa_{G}$, and $\lambda$ are from the references as given above. While, the reported value of $\sigma$ is found to be $10^{3}-10^{4}\, k_{B}T/\mu\textrm{m}$ for DOPC vesicle~\cite{KSGBPW} which is inconsistent with our finding value. The value of the obtained line tension $\sigma$ in our model seems to be determined by the balance between the stretching energy and the line tension energy only where the bending energy did not play a role. Somehow, our results support the study of pore dynamics by Brochard-Wyart \textit{et al.}~\cite{BWGS} in which the kinetic of a pore is driven by the two forces. Nevertheless, let us make some notes. 1) In their consideration, the Gaussian rigidity $\kappa_{G}$ did not appear, but we are. And the authors have assumed that the pore is much smaller than the whole vesicle which seems to be inconsistent with experimental observations. 2) The leak-out of the internal liquid has been considered in their model, while we did not investigate it. Experimental observations by Karatekin \textit{et al.}~\cite{KSGBPW} and Rodriguez \textit{et al.}~\cite{RCP} have been shown the leakage of the liquid through the pore during the open-closed states of the pore. This is presumably the origin of our discrepancy.

\subsection{Kinetics of birthing}
\label{subsec:kinetics}

\begin{figure}
\centering
\resizebox{0.6\textwidth}{!}{%
  \includegraphics{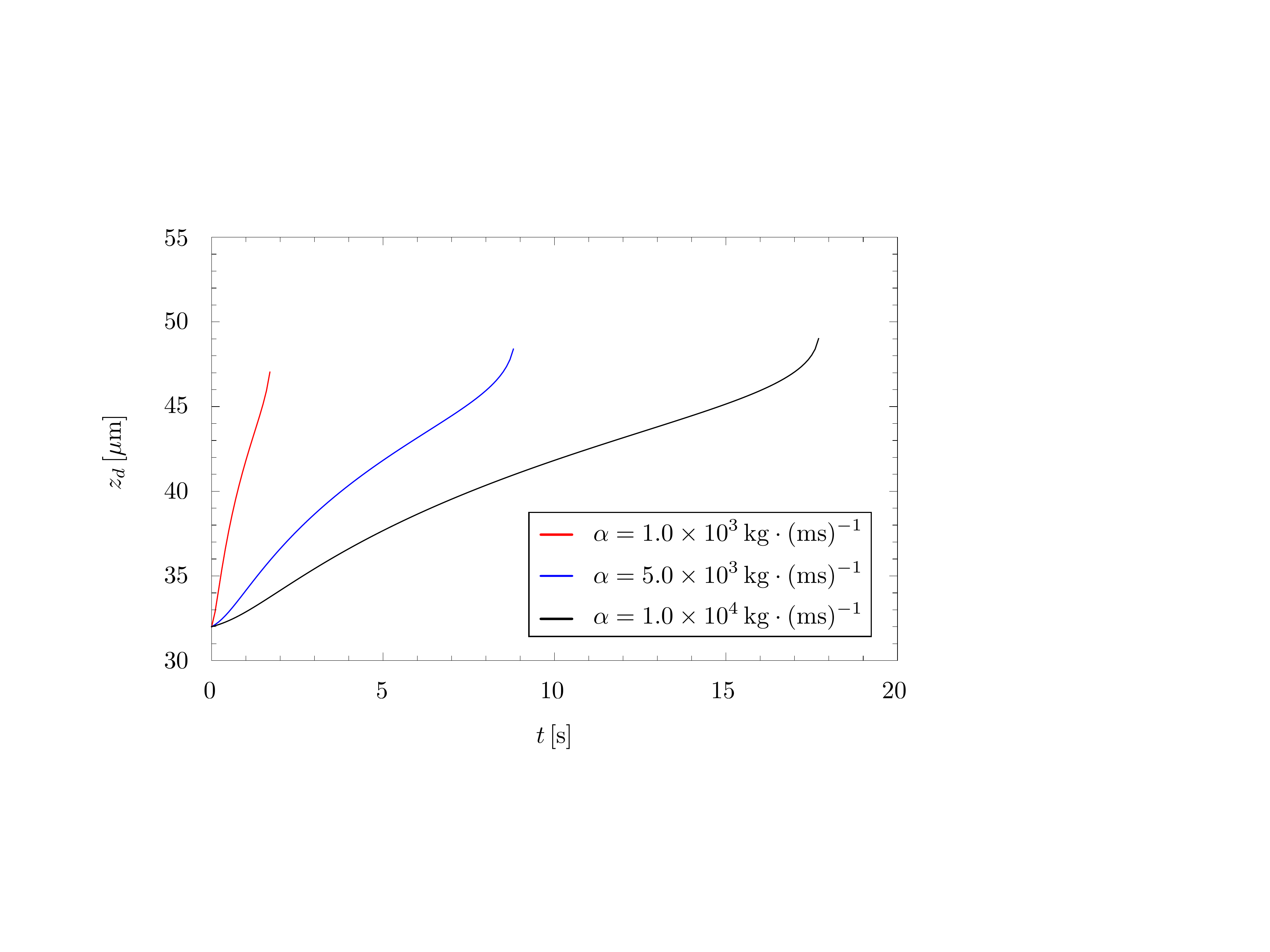}
}
\caption{The motion of the daughter vesicle with different values of the dynamic viscosity $\alpha$. The bending, stretching, and line tension moduli are fixed at $\kappa/\kappa_{G}=-1$, $\lambda=1.0\times 10^{8}\,  k_{B}T/\mu\textrm{m}^{2}$, and $\sigma=0.7\times 10^{8}\, k_{B}T/\mu\textrm{m}$, respectively. The results are evaluated at $T=35^{\circ}$C. $b=10\, \mu\textrm{m}$, $R_{\textrm{eq}}=38\, \mu\textrm{m}$, $R_{0}=40\, \mu\textrm{m}$, and $c_{0}=0$. The initial distance separating the two vesicles is $z_{d,0}=32\, \mu\textrm{m}$.}
\label{fig:friction_alpha}
\end{figure}

The free energy landscapes when $\kappa=20\, k_{B}T$, $\kappa_{G}=-20 \newline
k_{B}T$, $\lambda=1.0\times 10^{8}\,  k_{B}T/\mu\textrm{m}^{2}$, and $\sigma=0.7\times 10^{8}\, k_{B}T/\mu\textrm{m}$ are used to verify the kinetics of the birthing process at $T=35^{\circ}$C. By solving eq.~(\ref{equation_birthing}) numerically, we obtain the trajectory of the daughter vesicle when the dynamic viscosity $\alpha$ is changed as shown in fig.~\ref{fig:friction_alpha}. 
As the friction coefficient $\zeta$ decreases (by decreasing $\alpha$), the result shows clearly that the daughter vesicle has succeeded in the birthing with less translocation time. The daughter spends less distance to detach itself from the mother when its velocity increases as well. In fig.~\ref{fig:friction_alpha}, we have selected the values for $\alpha$ ranging from $1.0\times 10^{3}\, \textrm{kg}\cdot\textrm{(ms)}^{-1}-1.0\times 10^{4}\, \textrm{kg}\cdot\textrm{(ms)}^{-1}$. While, in our estimation, $\alpha$ is around $2.1\times 10^{3}\, \textrm{kg}\cdot\textrm{(ms)}^{-1}$ for the same values of $\lambda$ and $b$ (see Appendix A.). Figure~\ref{fig:friction_alpha} shows that $\alpha=1.0\times 10^{3}\, \textrm{kg}\cdot\textrm{(ms)}^{-1}$ yields the translocation time around 1.8 s which is close to the translocation time~\cite{YS} used in our estimation for $\alpha$. Note that our estimated $\alpha$ is directly obtained by knowing only the stretching modulus $\lambda$ with there is no information about $\kappa$, $\kappa_{G}$, and $\sigma$ included.

\begin{figure}
\centering
\resizebox{0.6\textwidth}{!}{%
  \includegraphics{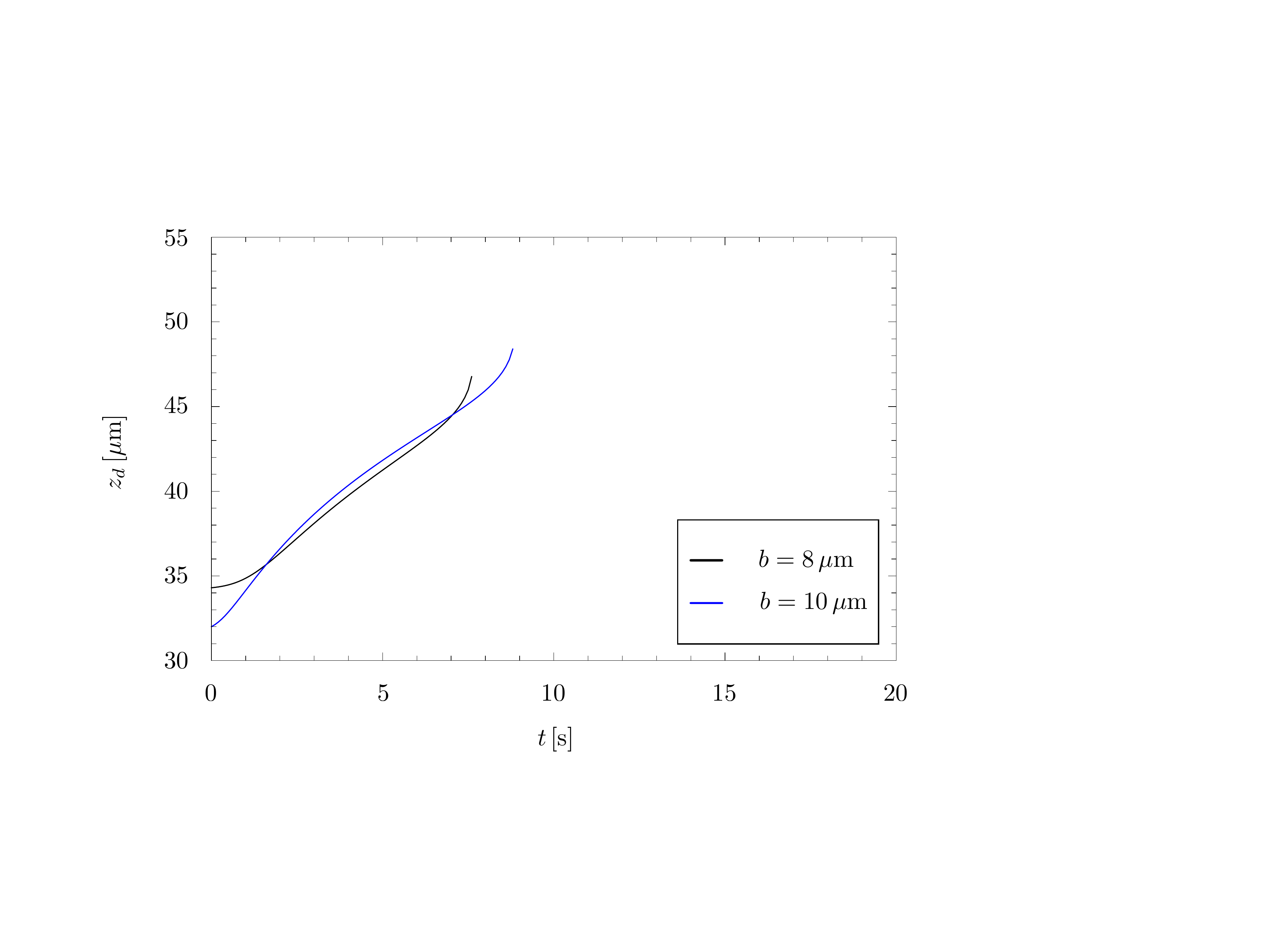}
}
\caption{The trajectory of the daughter vesicle with different size $b$. The bending, stretching, and line tension moduli are fixed at $\kappa/\kappa_{G}=-1$, $\lambda=1.0\times 10^{8}\,  k_{B}T/\mu\textrm{m}^{2}$, and $\sigma=0.7\times 10^{8}\, k_{B}T/\mu\textrm{m}$, respectively. The dynamic viscosity $\alpha$ is fixed at $5.0\times 10^{3}\, \textrm{kg}\cdot\textrm{(ms)}^{-1}$. The result is evaluated at $T=35^{\circ}$C. $R_{\textrm{eq}}=38\, \mu\textrm{m}$, $R_{0}=40\, \mu\textrm{m}$, and $c_{0}=0$.}
\label{fig:size}
\end{figure}

At $R_{\textrm{eq}}=38\, \mu\textrm{m}$ and $R_{0}=40\, \mu\textrm{m}$, fig.~\ref{fig:size} 
shows the trajectory of the daughter vesicle with different size $b$, when $\kappa/\kappa_{G}=-1$, $\lambda=1.0\times 10^{8}\,  k_{B}T/\mu\textrm{m}^{2}$, and $\sigma=0.7\times 10^{8}\, k_{B}T/\mu\textrm{m}$ are kept constant. The dynamic viscosity is fixed at $\alpha=5.0\times 10^{3}\, \textrm{kg}\cdot\textrm{(ms)}^{-1}$ and $T=35^{\circ}$C. The result indicates that, for the bigger daughter vesicle, the mother vesicle has more potential energy and, consequently, the mother tries to release the energy by expelling the daughter vesicle with a higher velocity than that of the smaller daughter vesicle in the initial stage. However, in the intermediate stage, the velocity is reduced due to the barrier from the line tension energy. The velocity of the daughter vesicle is growing up again due to the decrease of line energy and, finally,  the daughter vesicle is expelled completely from the mother vesicle. Also, the result shows that the smaller daughter takes less translocation time as expected.

\section{Conclusions}
 \label{sec:conclusions}

In this work, we have proposed a simple geometric ansatz to study the birthing of the daughter vesicle in the model system for self-reproducing vesicles, where the pore occurs on the membrane of the mother vesicle. To release the tension, the mother vesicle produces the pore on its surface and then the daughter is discharged from the mother. We have treated the deformation of the mother vesicle within the Helfrich free energy formalism. Our derived free energy suggests that changing the stretching and line tension moduli affects the free energy landscapes clearly, while changing the bending modulus does not affect the energy landscapes significantly. This suggests that the system is mainly governed by the competition between the stretching and line tension energies which support the previous theoretical~\cite{BWGS} and experimental~\cite{KSGBPW, RCP} works. Because the volume of the region between the mother and daughter vesicles is preserved (We assume no liquid flows out through the pore.), pressure does not play any roles in the model. The daughter vesicle is driven by the stretching energy which tries to shrink the mother vesicle's membrane. Onsager principle is used to derive the equation of motion for the birthing vesicle in terms of the bending, stretching and line tension moduli of the mother vesicle, and the size of the daughter vesicle, and the distance between the centers of the daughter and mother vesicles. We found that the translocation time decreases when the friction caused by the hole against the motion of the daughter vesicle decreases. Our results also suggest that when the daughter vesicle is large, the mother vesicle has more energy. Consequently, the larger daughter vesicle is initially expelled with a higher speed compared to the case of smaller daughter vesicle. However, the high speed will be decreased by the effect of barrier from the line tension energy in the later time. 

In our model, the specific details regarding the chemical properties of the vesicle are parametrized in terms of the material parameters such as bending, stretching, and line tension moduli for the purpose of investigating the large scale properties of the birthing process. The model suggests that the line tension modulus is of the order of ca. $10^{8}\, k_{B}T/\mu\textrm{m}$ for the successful birthing process, where we keep the bending and stretching moduli in accordance with the experimental reports. While, Karatekin \textit{et al.}~\cite{KSGBPW} reports that the value of the line tension is of the order of ca. $10^{3}-10^{4}\, k_{B}T/\mu\textrm{m}$ for single component membrane composed of cylinder-shaped lipid molecules (DOPC vesicle). Nevertheless, we would argue that our model is just a simple and first theoretical model to study the escape of a vesicle from a pore. In order to eliminate the discrepancy, we might need to consider the leakage of the liquid through the pore during the birthing process as observed by Karatekin \textit{et al.}~\cite{KSGBPW} and Rodriguez \textit{et al.}~\cite{RCP}. We will leave the improvement for our next work. However, in this work, we believe that our simple and handleable model captures the characteristic features of the birthing mechanism insightfully and it is a good starting point to study the vesicle discharged from the pore.
\newline
\newline
This work is supported by PCoMS-IPD program at Tohoku University and the Grant-in-Aid for Scientific Research (Grant Number 26287096 and 16K13844) from The Ministry of Education, Culture, Sports, Science and Technology (MEXT), Japan. 
%
%
%
%
%
\section{Authors contributions}
TK initiated this collaboration of the authors. PK performed research and calculations under the guidance of TK, and wrote the manuscript. YS and MI provided the experimental data. All the authors were involved in the preparation of the manuscript.
All the authors have read and approved the final manuscript.

\begin{appendices}

\renewcommand{\theequation}{A.\arabic{equation}}
\setcounter{equation}{0}
\section*{Appendix A. Estimation of the dynamic viscosity $\alpha$}
 \label{sec:alpha}

\begin{figure}
\centering
\resizebox{0.6\textwidth}{!}{%
  \includegraphics{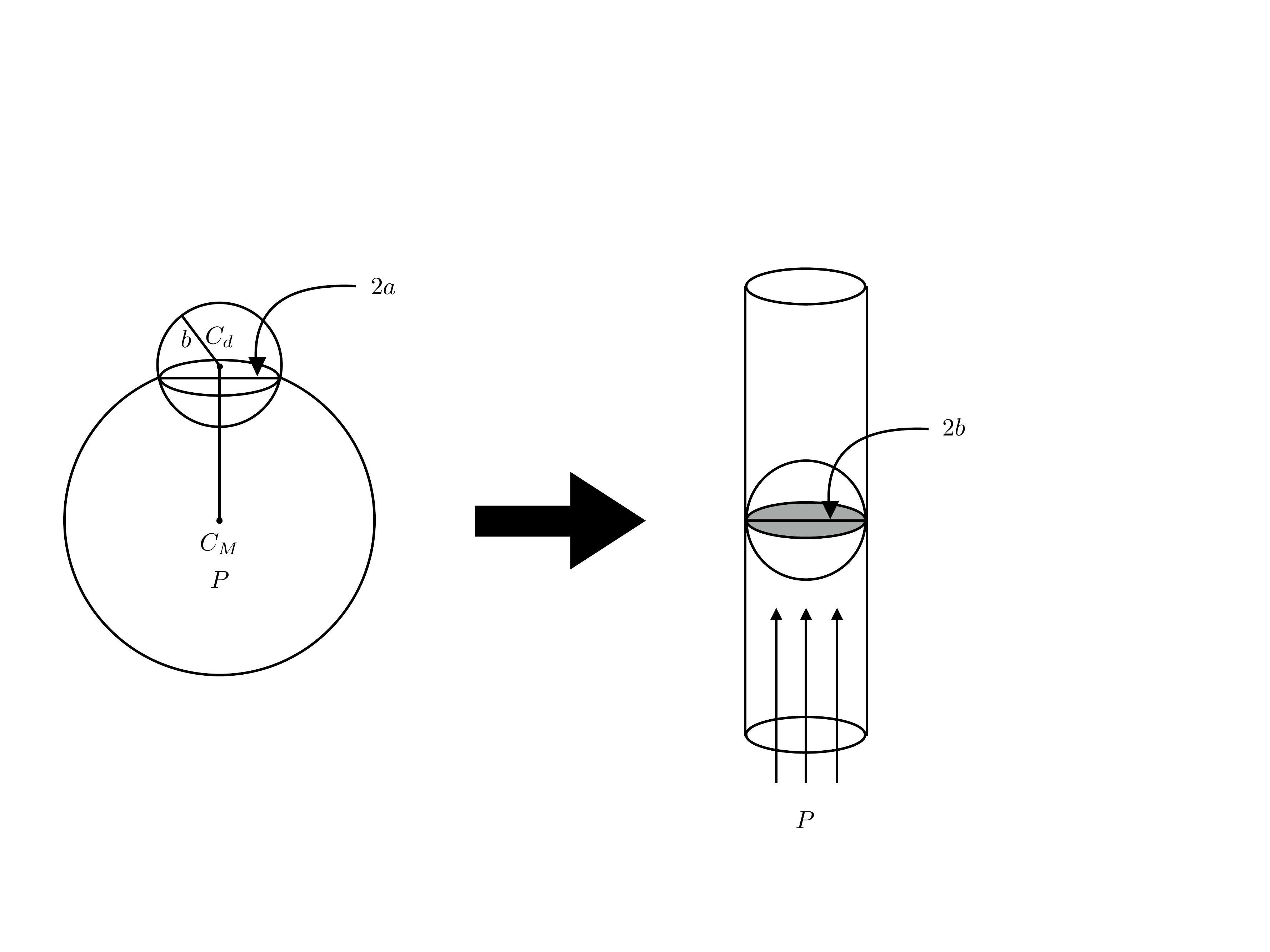}
}
\caption{Birthing of the vesicle passing through the pore is modelled as a rigid sphere flowing in the tube with a diameter $2b$.}
\label{fig:pressure model}
\end{figure}

In order to estimate the order of magnitude of $\alpha$, we have modelled the birthing of the daughter vesicle passing through the pore of radius $a$ by using a rigid sphere flowing through a tube with a diameter $2b$ as shown in fig.~\ref{fig:pressure model}. 

The pressure force acting on the sphere is
\begin{equation}
F=PA,
\end{equation}
where $P$ is the pressure difference between the lower and upper regions of the tube, and $A=\pi b^{2}$ is a crossection area of the tube.
Assuming the friction coefficient $\xi$ at the contact line, the steady-state velocity of the sphere $v$ is given by
\begin{equation}
F=\xi v,
\end{equation}
where the unit of $\xi$ is kg s$^{-1}$. Combining these two equations, we obtain
\begin{equation}
v=\frac{PA}{\xi}.
\end{equation}
The translocation time $t_{\textrm{trans}}$ is estimated by the time when the sphere migrates by a distance $2b$ with the speed $v$ as
\begin{equation}
\label{ttrans}
t_{\textrm{trans}}=\frac{2b}{v}=\frac{2b\xi}{PA}.
\end{equation}
Here, $\xi$ is given by
\begin{equation}
\label{xi}
\xi=\alpha L=2\pi\alpha b,
\end{equation}
where $\alpha$ has the unit of kg m$^{-1}$ s$^{-1}$. Substituting eq.~(\ref{xi}) and $A=\pi b^{2}$ into eq.~(\ref{ttrans}), we obtain 
\begin{equation}
\label{translocation}
t_{\textrm{trans}}=\frac{4\alpha}{P}.
\end{equation}
Then, we evaluate the pressure difference between the inner and outer regions of the vesicle, $P$. The stretching energy $F_{s}$ is defined as
\begin{equation}
\label{Fsapp}
F_{s}=\frac{\lambda}{2}\frac{(\Delta A)^{2}}{A_{\textrm{eq}}},
\end{equation}
where $A_{\textrm{eq}}$ is the equilibrium area of the vesicle and $\Delta A$ is the excess area due to the
stretching. Equation~(\ref{Fsapp}) can be rewritten as
\begin{equation}
\label{Fsa}
F_{s}=\frac{\lambda}{2}\frac{(\Delta A)^{2}}{A_{\textrm{eq}}}=A_{\textrm{eq}}\frac{\lambda}{2}\Bigg(\frac{\Delta A}{A_{\textrm{eq}}}\Bigg)^{2}=A_{\textrm{eq}}\frac{\lambda}{2}\Bigg(\frac{A-A_{\textrm{eq}}}{A_{\textrm{eq}}}\Bigg)^{2},
\end{equation}
where this expression means that the stretching energy is given by the square of the local strain $\Delta A/A_{\textrm{eq}}$ integrated over the whole vesicle surface which gives the factor $A_{\textrm{eq}}$.

Using the method of virtual work, this stretching energy $F_{s}$ can be given by an integral of the surface tension $\Sigma$ with respect to the excess area $\Delta A$ as
\begin{equation}
F_{s}=\int\Sigma d(\Delta A),
\end{equation}
and, therefore,
\begin{equation}
\label{surface tension}
\Sigma=\frac{\partial F_{s}}{\partial(\Delta A)}=\lambda\frac{\Delta A}{A_{\textrm{eq}}}.
\end{equation}
Then, the pressure difference $P$ can be given by Laplace law as
\begin{equation}
\label{pressure}
P=\frac{2\Sigma}{b}=\frac{2\lambda}{b}\frac{\Delta A}{A_{\textrm{eq}}}.
\end{equation}
Substituting eq.~(\ref{pressure}) into eq.~(\ref{translocation}), we obtain
\begin{equation}
t_{\textrm{trans}}=\frac{2\alpha b}{\lambda}\frac{A_{\textrm{eq}}}{\Delta A}.
\end{equation}
Thus,
\begin{equation}
\label{alphab}
\frac{\alpha b}{\lambda}=\frac{t_{\textrm{trans}}}{2}\frac{\Delta A}{A_{\textrm{eq}}}.
\end{equation}
Experiments show that $t_{\textrm{trans}}\approx1.0\, \textrm{s}$~\cite{YS} and $\Delta A/A_{\textrm{eq}}=0.1$. Substituting $b=10\, \mu\textrm{m}$ and $\lambda=1.0\times 10^{8}\, k_{B}T/\mu\textrm{m}^{2}$ into eq.~(\ref{alphab}), we, then, obtain 
\begin{equation}
\alpha\approx2.1\times10^{3}\, \textrm{kg m}^{-1}\textrm{s}^{-1}. 
\end{equation}

\end{appendices}

%

\begin{thebibliography}{}
\bibitem{YS} Y.~Sakuma, The resealing time of the pore (translocation time of detached daughter vesicle) is measured in the developed model system for self-reproduction DLPE/DPPC vesicle. Since the pore opens and completely closes, the measure indicates that the translocation time is around a second with the diameter of the mother and daughter vesicles are 28.2$\, \mu\textrm{m}$ and 13.1$\, \mu\textrm{m}$, respectively.
\end{thebibliography}
%

\end{document}